# Correcting User Decisions Based on Incorrect Machine Learning Decisions

Saveli Goldberg[1], Lev Salnikov[2], Noor Kaiser[3], Tushar Srivastava[3], and Eugene Pinsky[3]

[1] Department of Radiation Oncology Mass General Hospital, Boston, MA 02115, USA emailsigoldberg@mgh.harvard.edu

[2] AntiCA Biomed, San Diego, CA emaillsalnikov@anticabiomed.com

[3] Department of Computer Science, Metropolitan College, Boston University 1010 Commonwealth Avenue, Boston, MA 2215, USA nkaiser@bu.edu, tushar98@bu.edu, epinsky@bu.edu

**Abstract.** It is typically assumed that for the successful use of machine learning algorithms, these algorithms should have a higher accuracy than a human expert. Moreover, if the average accuracy of ML algorithms is lower than that of a human expert, such algorithms should not be considered and are counter-productive. However, this is not always true. We provide strong statistical evidence that shows that even if a human expert is more accurate than a machine, an interaction with such a machine is beneficial when communication with the machine is non-public. The existence of a conflict between the user and ML model, and the private nature of user-AI communication will have the effect of making the user think about their decision and hence increase overall accuracy.

## 1 Introduction

As the use of machine learning (ML) and AI becomes ever more widespread, it is inevitable that the decisions by human experts and those by ML systems would inevitably differ in numerous situations. How should such conflicts be resolved?

Intuitively, if the results of ML algorithms are more accurate than the decisions of an expert, this would help the expert to make a well informed, and correct decision. On the flip-side, when the results of ML algorithms are lower in accuracy than the decisions of an expert, the ML decisions are expected to be counter-productive. However, our experiments suggest that even in cases where the ML model's accuracy is lower than that of an expert, such ML decisions are still beneficial and useful in the decision making process for an expert – as long as the ML model's accuracy is not significantly lower than that of an expert.

How can we explain such a surprising result? We show that in case of such a conflict and when the ML model's decisions are not publicly known, an expert would think more critically about their decision and subsequently achieve higher accuracy. We illustrate this by running a series of experiments.

## 2     Experiment Design

In order to design our experiments to accurately replicate a situation where an expert consults an oracle, we considered experimenting with music since many people have expertise in this area. Additionally, we had several participants willing to partake in these experiments because the subject matter of the tests was interesting and familiar to them, and because they were eager to learn more about the research we were conducting. It is of value to note that all participants had differing levels of knowledge in this realm, and that decisions of the participants had no consequences.

There were 2 types of tests the participants of this study completed in all rounds, across time. 34 different songs were used to create these tests. Tests 1 and 2 consisted of **12 questions**, and Test 3 consisted of **10 questions**.

The pattern of these tests was as follows: first, participants are presented with an audio clip and are asked to choose who they believe the artist of the song is (refer to **Figure 1** below for an example of this question). Once they make their preliminary decision, they move on to the next question where the same audio clip is played again and the artist identified by the machine learning model is displayed, too. They are then asked to make their final decision for the artist of the song (refer to **Figure 2** below for an example of this question). This gives them the opportunity to either choose to stick with their preliminary response, or change their final response anonymously in light of the new information presented to them **after** choosing their preliminary response.

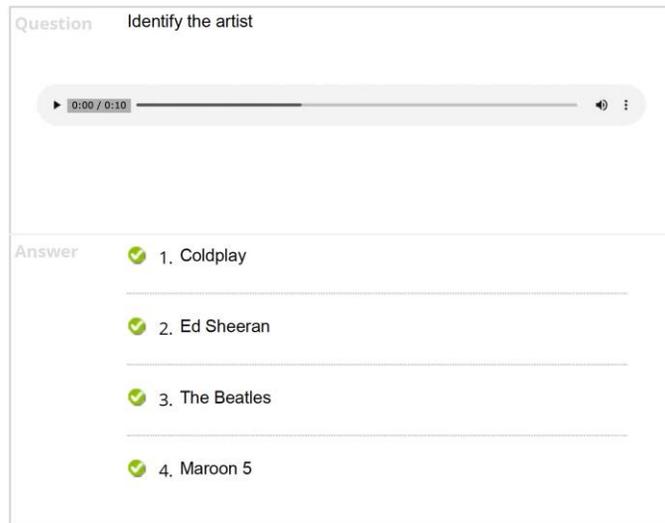

**Fig. 1.** Preliminary decision question

Both types of tests were completed independently of one another. In order to standardise all the tests, the audio clips presented to the participants were 10 seconds in length each, and the time given to complete each test was set to 10 minutes. It is important to note that backtracking was not permitted in these tests, i.e., once participants had recorded a response and moved on to the next question, they were not able to go back to their previous questions to view or change their responses. This ensured that the integrity of the data collected would remain intact. Additionally, participants were not allowed to move to the next question without responding to the current question, which ensured that the data collected was complete since the likelihood of missing information or incomplete fields was eliminated.

**Fig. 2.** Final decision question

Three levels of accuracy of the machine learning model were used; **66.67%, 75%** and **80%**. For example, when the accuracy of the machine learning model is set to 80% in the 3rd test, 8 out of 10 of the *"system responses"* shown to the participants in the test are correct, while 2 out of 10 of these responses are incorrect. The participant is unaware of what the accuracy level of the ML model is for their respective tests.

All participants completed these tests on Blackboard Learn, which is the primary learning management system at Boston University. Participants were able to access and complete these tests on this platform at any time within a 3 day period once the tests were made available to them as part of their course. However, as mentioned earlier, the performance on these tests was not treated as part of their coursework, i.e., participants were not rewarded or penalised for their performance on these tests in any manner.

# 3   Results

A general description of the experimental results is presented in **Table 1**, which shows that the final solution of the students turned out to be much more accurate than both the students' preliminary solution and the ML solution.

Additionally, it should be noted that in each of the three experiments, the preliminary accuracy of the participants was higher than the accuracy of the ML model. A detailed distribution of test results for each class of students is presented in **Tables 2.1, 2.2** and **2.3**.

**Table 1.** Summary Statistics for all Student Classes

| ML Acc | No. of Songs | Prelim Acc $\mu$ | $\sigma$ | Final Acc $\mu$ | $\sigma$ | No. of Students | ML vs. Prelim p-value | ML vs. Final p-value | Prelim vs. Final p-value |
|--------|--------------|------------------|----------|-----------------|----------|-----------------|----------------------|---------------------|--------------------------|
| 66.7% | 12 | 77.5% | 24.5% | 83.1% | 17% | 239 | < 0.0001 | < 0.0001 | < 0.0001 |
| 75% | 12 | 81% | 25.1% | 87.6% | 18% | 228 | < 0.0001 | < 0.0001 | < 0.0001 |
| 80% | 10 | 85.4% | 21.1% | 90.7% | 12.6% | 173 | 0.0009 | < 0.0001 | < 0.0001 |

**Table 2.1.** Statistics by Student Classes

| Machine Accurary = 66.7% | | | | | | | |
|---|---|---|---|---|---|---|---|
| Class | Year | Prelim Acc $\mu$ | $\sigma$ | Final Acc $\mu$ | $\sigma$ | No. of Students | Prelim vs. Final p-value |
| CS 521 A | 1 | 76.4% | 24.5% | 80.5% | 20.1% | 29 | 0.3357 |
| CS 521 B | 2 | 74.0 % | 27.7% | 78.5% | 20.2% | 33 | 0.1040 |
| CS 550 A | 1 | 78.6% | 25.6% | 82.3% | 12.5% | 16 | 0.4600 |
| CS 550 B | 2 | 80.4% | 20.4% | 83.9% | 15.7% | 31 | 0.0965 |
| CS 677 A | 1 | 80.5% | 23.5% | 86.0% | 14.8% | 38 | 0.0084 |
| CS 677 B | 2 | 79.6% | 22.2% | 84.8% | 15.6% | 29 | 0.0287 |
| CS 677 C | 2 | 79.2% | 25.5% | 84.3% | 16.7% | 26 | 0.0614 |
| CS 677 D | 1 | 73.0% | 27.2% | 83.6% | 17.6% | 37 | 0.0004 |

**Table 2.2.** Statistics by Student Classes (continued)

| Machine Accuracy = 75% | | | | | | | |
|---|---|---|---|---|---|---|---|
| Class | Year | Prelim Acc $\mu$ | $\sigma$ | Final Acc $\mu$ | $\sigma$ | No. of Students | Prelim vs. Final p-value |
| CS 521 A | 1 | 79.8% | 22.1% | 88.9% | 12.7% | 21 | 0.0098 |
| CS 521 B | 2 | 75.8% | 23.2% | 83.8% | 14.6% | 33 | 0.0016 |
| CS 550 A | 1 | 83.3% | 16.4% | 86.1% | 9.8% | 15 | 0.3712 |
| CS 550 B | 2 | 82.0% | 18.3% | 85.5% | 12.5% | 31 | 0.1191 |
| CS 677 A | 1 | 83.8% | 20.6% | 90.4% | 14.6% | 38 | 0.0004 |
| CS 677 B | 2 | 83.6% | 19.9% | 88.8% | 12.3% | 29 | 0.0099 |
| CS 677 C | 2 | 82.3% | 21.0% | 88.7% | 13.1% | 25 | 0.0117 |
| CS 677 D | 1 | 78.7% | 22.7% | 88.0% | 14.1% | 36 | 0.0007 |

**Table 2.3.** Statistics by Student Classes (continued)

| Machine Accuracy = 80% | | | | | | | |
|---|---|---|---|---|---|---|---|
| Class | Year | Prelim Acc $\mu$ | $\sigma$ | Final Acc $\mu$ | $\sigma$ | No. of Students | Prelim vs. Final p-value |
| CS 521 A | 1 | 89.1% | 12.2% | 90.9% | 15.1% | 11 | 0.5059 |
| CS 521 B | 2 | 83.7% | 26.2% | 88.3% | 20.4% | 30 | 0.0996 |
| CS 550 A | 1 | 89.3% | 15.9% | 91.4% | 8.6% | 14 | 0.4257 |
| CS 550 B | 2 | 91.9% | 12.1% | 92.2% | 9.7% | 27 | 0.8320 |
| CS 677 A | 1 | 79.7% | 24.3% | 89.7% | 9.4% | 38 | 0.0035 |
| CS 677 B | 2 | 90.7% | 14.1% | 95.2% | 7.0% | 27 | 0.0310 |
| CS 677 C | 2 | 80.0% | 26.1% | 88.1% | 12.7% | 26 | 0.0480 |

A more complex understanding of the experiments is captured in **Tables 3.1, 3.2** and **3.3**, which present the results of the final conclusion for each subgroup of students with the same preliminary solution accuracy.

**Table 3.1.** Preliminary Result Statistics

| Machine Accuracy = 66.7% | | | | |
|---|---|---|---|---|
| Prelim Acc Interval | Final Acc $\mu$ | $\sigma$ | No. of Students | p-value |
| 0% | 50% | – | 1 | – |
| 16.7% | 59.3% | 12.1% | 9 | 0.0001 |
| 25% | 50% | 16.7% | 3 | – |
| 33.3% | 60% | 11.7% | 10 | 0.0001 |
| 41.7% | 58.3% | 23.6% | 2 | – |
| 50% | 66.7% | 13% | 24 | 0.0001 |
| 58.3% | 75% | 5.6% | 10 | 0.0001 |
| 66.7% | 69.7% | 13.5% | 22 | 0.3048 |
| 75% | 84.4% | 7.9% | 16 | 0.0003 |
| 83.3% | 82.3% | 12.6% | 33 | 0.6501 |
| 91.7% | 91.7% | 4.8% | 25 | 0.9975 |
| 100% | 97.1% | 7.6% | 84 | 0.0008 |

**Table 3.2.** Preliminary Result Statistics (continued)

| Machine Accuracy = 75% | | | | |
|---|---|---|---|---|
| Prelim Acc Interval | Final Acc $\mu$ | $\sigma$ | No. of Students | p-value |
| 16.67% | 69.4% | 9.6% | 3 | – |
| 25% | 37.5% | 5.9% | 2 | – |
| 33.3% | 66.7% | 9.1% | 6 | 0.0003 |
| 41.7% | 69.4% | 12.7% | 3 | – |
| 50% | 70.1% | 13.5% | 12 | 0.0067 |
| 58.3% | 75% | 9.9% | 21 | 0.0001 |
| 66.7% | 80.6% | 6.6% | 21 | 0.0001 |
| 75% | 81.7% | 10.7% | 20 | 0.0116 |
| 83.3% | 87.5% | 6% | 24 | 0.0025 |
| 91.7% | 92.9% | 5.1% | 34 | 0.17 |
| 100% | 98.5% | 4.6% | 82 | 0.0033 |

**Table 3.3.** Preliminary Result Statistics (continued)

| Machine Accuracy = 80% | | | | |
|---|---|---|---|---|
| Prelim Acc Interval | Final Acc $\mu$ | $\sigma$ | No. of Students | p-value |
| 10% | 80% | – | 1 | – |
| 20% | 65% | 30% | 4 | – |
| 30% | 90% | – | 1 | – |
| 40% | 72.9% | 18.9% | 7 | 0.0037 |
| 50% | 72.9% | 11.1% | 7 | 0.0036 |
| 60% | 73.3% | 12.1% | 6 | 0.0430 |
| 70% | 82% | 11.5% | 15 | 0.0012 |
| 80% | 90% | 4.5% | 11 | 0.0001 |
| 90% | 90.9% | 6.9% | 32 | 0.4509 |
| 100% | 97.8% | 4.7% | 88 | 0.0001 |

**Table 4.** Statistics by Preliminary Decision Intervals

| Machine Accuracy = 66.7% | | | | | | | |
|---|---|---|---|---|---|---|---|
| Interval of Preliminary Decision | Prelim Acc $\mu$ | $\sigma$ | Final Acc $\mu$ | $\sigma$ | No. of Students | Prelim vs. Final p-value | ML vs. Final p-value |
| 56%-76% | 67.67% | 6.22% | 75.67% | 12% | 50 | < 0.0001 | < 0.0001 |
| < 56% | 37.59% | 14.24% | 62.24% | 13.4% | 49 | < 0.0001 | 0.0129 |
| > 76% | 94.71% | 6.95% | 92.29% | 10.83% | 148 | < 0.0001 | < 0.0001 |
| Machine Accuracy = 75% | | | | | | | |
| Interval of Preliminary Decision | Prelim Acc $\mu$ | $\sigma$ | Final Acc $\mu$ | $\sigma$ | No. of Students | Prelim vs. Final p-value | ML vs. Final p-value |
| 64%-86% | 75.38% | 6.97% | 83.46% | 8.4% | 65 | < 0.0001 | < 0.0001 |
| < 64% | 47.87% | 12.94% | 70.39% | 12.86% | 47 | < 0.0001 | 0.0172 |
| > 86% | 97.56% | 3.81% | 96.84% | 5.35% | 116 | 0.1142 | < 0.0001 |
| Machine Accuracy = 80% | | | | | | | |
| Interval of Preliminary Decision | Prelim Acc $\mu$ | $\sigma$ | Final Acc $\mu$ | $\sigma$ | No. of Students | Prelim vs. Final p-value | ML vs. Final p-value |
| 68%-92% | 82.93% | 8.59% | 88.45% | 8.75% | 58 | < 0.0001 | < 0.0001 |
| < 68% | 43.33% | 14.94% | 72.22% | 16.49% | 27 | < 0.0001 | < 0.0001 |
| > 92% | 100% | – | 97.84% | 4.66% | 88 | < 0.0001 | < 0.0001 |

Finally, **Table 4** summarizes these results for three groups of students as follows:

- – Group 1 – students with a preliminary decision in the range from 85% ML solution to 115% ML solution
- – Group 2 – students with a preliminary solution of less than 85% of the ML solution
- – Group 3 – students with a preliminary solution of more than 115% of the ML solution

## 4   Discussion

At first glance, our results seem counter-intuitive. Why would an expert consult a machine learning system with lower accuracy? The reason for this is the following: if an expert's opinion differs from that of a machine, the expert would think more about their decision. This was enough to produce a measurable effect. Moreover, in arriving at a final decision, many important negative factors arising from disputes between people are ignored (such as the a priori authority of expert ambitions of participants).

With this in mind, we need to note the important feature of our experiments: quality of the decisions **does** not have the consequences for the expert. In cases where a wrong decision has been documented and could have affected the expert's professional standing or career, the results of the tests could have been different. As a result, the absence of such documented results by machine learning could benefit human - AI interaction.

As we can see, if the preliminary accuracy of a student was much lower than the accuracy of the ML model, then the final decision was better than a preliminary one but still not higher than that of the ML model. On the other hand, if the preliminary accuracy of a student was much higher than that of the ML model, then the final accuracy was lower than the preliminary decision but still better than that of the ML model. We can see this when the preliminary accuracy was 100%.

The greatest interest is the group of students with accuracy close to that of the ML model, specifically with ML accuracy ±15% **of ML accuracy** (56%-76% for 66.7%, 64%-86% for 75% and 68%-92% for 80%). In this case, the accuracy of the final decision was much higher than the preliminary decision and the ML model's decision (paired T-test $p < 0.0001$)**:**

- – 80%        ML accuracy: 82.9% ±−1.1% preliminary decision and 88.4% ±−1.1% final decision
- – 66.7% ML accuracy: 67.7% ± 0.9% preliminary decision and 75.7% ± 1.7% final decision
- – 75%        ML accuracy: 75.4% ± 0.9% preliminary decision and 83.5% ± 1.0% final decision

Examining the tables, we see the following:

1. If we take students with preliminary accuracy less than 56% when ML had 66.7%, the average preliminary is 37.6% +/- 2% and final 62.2 +/- 1.9%

2. If we take students with preliminary accuracy less than 64% when ML had 7**5**%, the average preliminary is 47.9% +/- 1.9% and final 70.4% +/- 1.9%

3. If we take students with preliminary accuracy less than 68% when ML had 80%, the average preliminary is 43.3% +/- 2.9% and final 72.2 +/- 3.2%

4. If we took students with preliminary accuracy of more than 76% when ML had 66.7%, the average preliminary is 94.7% +/- 0.6% and final 92.3 +/- 0.9%

5. If we took students with preliminary accuracy of more than 86% when ML had 75%, the average preliminary is 97.6% +/- 0.4% and final 96.8% +/- 0.5%

6. If we took students with preliminary accuracy of more than 92% when ML had 80%, preliminary is 100%, and final 97.8 +/- 0.5%

## 5    Summary and Conclusion

We demonstrated that the human-AI interaction can be beneficial even in cases where an ML model has a lower accuracy than an expert – provided that the ML model's accuracy is not significantly lower than the expert's. **An important condition for this was that the discrepancies between the decisions of the ML and the subject did not affect the fate of the subject. In real life, this is equivalent to the fact** that the ML decision is known only to the expert and is not documented outside of this dialog, i.e., the interaction between the human and ML model is private and anonymous. It is possible that if the anonymity in consulting an oracle to inform or change your decision was not private, our results would be different.

In the future, we intend on extending our work in other applications as follows: we aim to use other media (such as images, text etc.) from various domains or subject matter other than music to attempt to establish conclusions similar to those we reached after this set of experiments.

Additionally, an interesting idea we intend on investigating is the impact of incentives on our results. How would our results change if participants were offered a reward for making a correct decision? Will the results we obtain be more pronounced than those we got when we experimented without rewards? We hypothesize that the performance of a participant will be closer to that of the ML model's accuracy since it is likely that the participants give more credence to the "system responses" presented to them during the experiments. Our aim is to redesign our existing experiments in order to prove our hypothesis.